\pdfoutput=1
\documentclass[twocolumn,showpacs,prl,superscriptaddress]{revtex4-1}
\usepackage{graphicx}
\usepackage{amsfonts}
\usepackage{amsmath} 
\usepackage{wasysym} 
\usepackage{amssymb} 
\usepackage{enumitem} 
\usepackage[english]{babel} 
\usepackage{color} 

\graphicspath{{figures/}}

\usepackage[]{hyperref}
\hypersetup{
  colorlinks   = true, 
  urlcolor     = blue, 
  linkcolor    = blue, 
  citecolor   = red 
}

\newcommand{\segl}{r}

\newcommand{\ensavp}[2]{\langle{#1(t;#2)}\rangle}
\newcommand{\ensavpr}[1]{\widetilde{\langle{#1(t)}\rangle}}
\newcommand{\densone}[2]{u_{#2}(#1)}

\newcommand{\densavpos}[1]{\overline{u}(#1)}
\newcommand{\ay}{|y|}

\newcommand{\zJ}[1]{j_{#1}}

\newcommand{\J}{\ensuremath{\operatorname{J}}}
\newcommand{\nsumbess}[2]{\zeta_{#1}({#2})}
\newcommand{\bsumbess}[3]{\sum_{n=1}^{\infty} \frac{1}{\J_{#2}(\zJ{#1,n})\zJ{#1,n}^{#3}}}
\newcommand{\nsumJ}[2]{\nsumbess{\J\!{#1}}{#2}}
\newcommand{\nsumJnu}[1]{\nsumJ{\nu}{#1}}

\newcommand{\njsum}[1]{\nsumbess{\J\! 1}{#1}}

\renewcommand{\d}[1]{\ensuremath{\operatorname{d}\!{#1}}}
\newcommand{\tpxh}[1]{{2\pi[{#1}+1/2}]}
\newcommand{\tpnh}{\tpxh{n}}

\newcommand{\ein}[1]{\ensuremath\operatorname{Ein}({#1})}

\newcommand{\einbig}[1]{\ensuremath\operatorname{Ein}\!\left({#1}\right)}
\newcommand{\plr}{r}

\newcommand{\myheading}[1]{{\textit{#1}---}}

\begin{document}
\title{Anomalous diffusion from Brownian motion with random confinement}
\newcommand{\icfoaf}{\affiliation{ICFO--Institut de Ci\`encies Fot\`oniques, Mediterranean
Technology Park, 08860 Castelldefels, Spain}}
\newcommand{\csicaf}{\affiliation{Spanish National Research Council (IDAEA-CSIC), E-08034 Barcelona, Spain}}
\author{G. J. \surname{Lapeyre}, Jr.}
\csicaf
\icfoaf
\date{\today}

\begin{abstract}
  We present a model of anomalous diffusion consisting of an ensemble
  of particles undergoing homogeneous Brownian motion except for
  confinement by randomly placed reflecting boundaries. For power-law
  distributed compartment sizes, we calculate exact and asymptotic
  values of the ensemble averaged mean squared displacement and find
  that it increases subdiffusively, as either a power or the logarithm of
  time. Numerical simulations show that the probability density function of
  the displacement is non-Gaussian. We discuss the relevance of the model
  for the analysis of single-particle tracking experiments and its relation
  to other sources of subdiffusion. In particular we discuss an intimate connection
  with diffusion on percolation processes.
\end{abstract}


\maketitle


In the study of diffusion in physical
systems it is well known that a variety of processes lead to diffusion that
deviates from pure Brownian motion~\cite{Havlin1987,Bouchaud1990,Metzler2004,Barkai2012a,Hoefling2013,Metzler2014}.
The most commonly studied signature of anomalous diffusion
is the mean squared displacement (MSD), which often takes the form of a power law,
\begin{equation}
 \label{mainmsd}
  \langle x^2(t) \rangle \sim t^\beta,
\end{equation}
where $x(t)$ is the displacement of a particle at time
$t$. For Brownian motion $\beta=1$, and for subdiffusive motion $\beta<1$.
Subdiffusive systems may be successfully modeled by various random
processes such as continuous time random walk
(CTRW)~\cite{ScherMontroll75,Klafter1987,Klafter2011}, fractional
Brownian motion~\cite{Kolmogorov1940,Mandelbrot1968} (FBM), diffusion
on percolation~\cite{Stauffer1991} (DOP), Lorentz
models~\cite{Hoefling2013}, or by a combination of
these~\cite{Meroz2010,Weigel2011}. The problem of distinguishing which
processes contribute to or modify anomalous diffusion is an active
area of
research~\cite{Magdziarz2009,Magdziarz2011,Jeon2013,Meroz2013}.  In
particular, one may need to model complicating factors such as
confinement. Confinement plays an important role in modifying or
attenuating the subdiffusion manifested by the power-law behavior
in~\eqref{mainmsd}~\cite{Meilhac2006,Condamin2007,Destainville2008,Condamin2008,Denisov2008,Burada2009,Neusius2009,Magdziarz2010,Jeon2012,Bruna2014}.  But, in this paper we take a
different view of the relation of confinement to subdiffusion: We show
that random confinement may in fact be the sole cause of observed
subdiffusion. While the effect is quite general, the model and
analysis are motivated specifically by single particle tracking (SPT)
experiments, in particular studies of the biophysics of
cells~\cite{Tolic04,Golding06,Burov2011,Giannone13}. We present a
scenario in which disordered confinement produces quantitative
signatures of subdiffusion in a typical analysis of SPT data. Given
that heterogeneous confining boundaries are often observed in
biophysics, it follows that in experiments searching for contributions
to the subdiffusion exponent $\beta$, confinement may not be discarded
\textit{a priori} as a candidate.

\myheading{Experimental scenario} Consider a number of Brownian
particles uniformly distributed in a space that is partitioned into
compartments by a random arrangement of reflecting barriers.
Apart from the presence of the boundaries, the motion is diffusive,
with parameters homogeneous is space and time.  We collect an
SPT trajectory (time series) for each of several particles sampled
uniformly from the space. The trajectory consists of the displacement
of each particle from its starting point recorded at a series of
times.  Typically, the trajectories are first analyzed via the 
ensemble averaged mean squared displacement (EMSD),
or the time-ensemble averaged mean squared displacement (TEMSD), where the
TEMSD consists, operationally, of first computing a time (sliding) averaged MSD
 (TMSD) for each trajectory and then averaging the result over the trajectories.
For the minimal model introduced below, we find 
\begin{itemize}[noitemsep,topsep=0pt]
\setlength\itemsep{0em}
\item The EMSD is unbounded and subdiffusive in the sense of \eqref{mainmsd} if 
 the distribution of the linear size of the compartments has a sufficiently heavy-tail.
\item There is no weak ergodicity breaking (if the system is in equilibrium): the EMSD is equal to the TEMSD.
\item The TMSD (ie for a single trajectory) tends to a constant.
That is, the observables of a trajectory are not self-averaging over disorder.
\end{itemize}

\begin{figure}[t!]
   \includegraphics[width= \columnwidth]{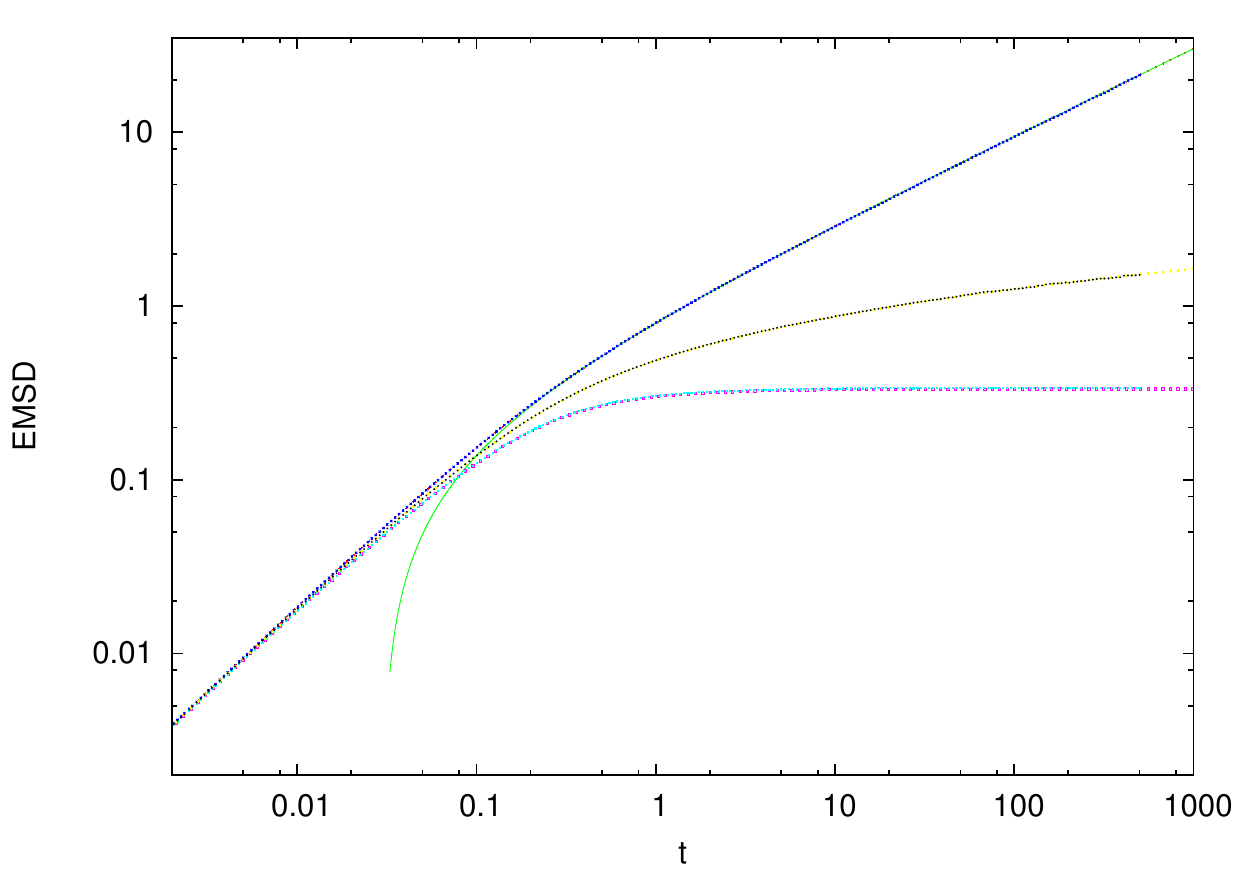}
   \caption{
   EMSD for random confinement model RSFC with $r_0 = 1$ showing
   bounded  motion ($\alpha =4$, lower curve), power-law ($\alpha=1$, upper curve), and
  logarithmic ($\alpha=2$, middle curve) growth. Each of the three curves has points from 
   exact solutions (\eqref{genone} and \eqref{logsol}) and simulations.
   Green curve is asymptotic solution \eqref{onedsolone} for $\alpha=1$.
  Simulations are
  a rescaled random walk with $r_0=100$ in lattice units. 
  Averages are over $1.8 \times 10^8$ trajectories, and $10^6$ steps per
  trajectory, using the Mersenne Twister random number generator.
}
\label{fig:msdcurves}
\end{figure}

\myheading{The model} We choose a minimal model that captures the
essential features and displays asymptotically subdiffusive motion.
We refer to the particular model presented below as the random
scale-free confinement model (RSFC).  We state the results for the MSD
before giving the detailed calculations.  In this paper, we treat in
detail only one spatial dimension, stating some results for higher
dimension at the end, but leaving details to a subsequent paper.  The
probability density for the displacement of a particle diffusing on a
line segment of length $\segl$ with reflecting boundary conditions
obeys
\begin{equation}
 \label{diffeqn}
 \partial_tu(x;t) = \partial^2_x u(x;t),
\end{equation}
with $\partial_x u(\segl/2;t) = \partial_x u(-\segl/2;t) = 0$.
We assume that the probability for a particle to be found in a segment of length $\segl$ is 
Pareto-distributed
\begin{equation}
\label{pareto}
  P(\plr) =  P_\alpha(\plr) = \begin{cases}
    0 \quad \text{ for } \plr \le r_0 \\
    \alpha r_0^{\alpha}\plr^{-\alpha-1} \text{ for } \plr > r_0
  \end{cases},  \text{for } \alpha > 0.
\end{equation}
We find that the asymptotic EMSD is, for $\alpha\ne 2$
\begin{equation}
\label{genpower}
  \ensavpr{x^2} = K_1 + K_2 t^{\frac{2-\alpha}{2}},
\end{equation}
so that the EMSD either decays to a constant for $\alpha >2$ or grows without bound
for $\alpha < 2$. For $\alpha = 2$ we find
\begin{equation}
\label{genlog}
  \ensavpr{x^2} = K_3 + K_4 \ln(t).
\end{equation}
In \eqref{genpower} and \eqref{genlog} angle brackets denote averaging over particle trajectories,
the tilde denotes averaging over the disorder (ie over $\plr$), and $K_1,K_2,K_3,K_4$ are constants that depend on dimension,
and the initial distribution of particles.
Examples of curves for three values of $\alpha$ showing both bounded and unbounded EMSD are shown in Fig.~\ref{fig:msdcurves}.

\begin{figure}[t!]
   \includegraphics[width= \columnwidth]{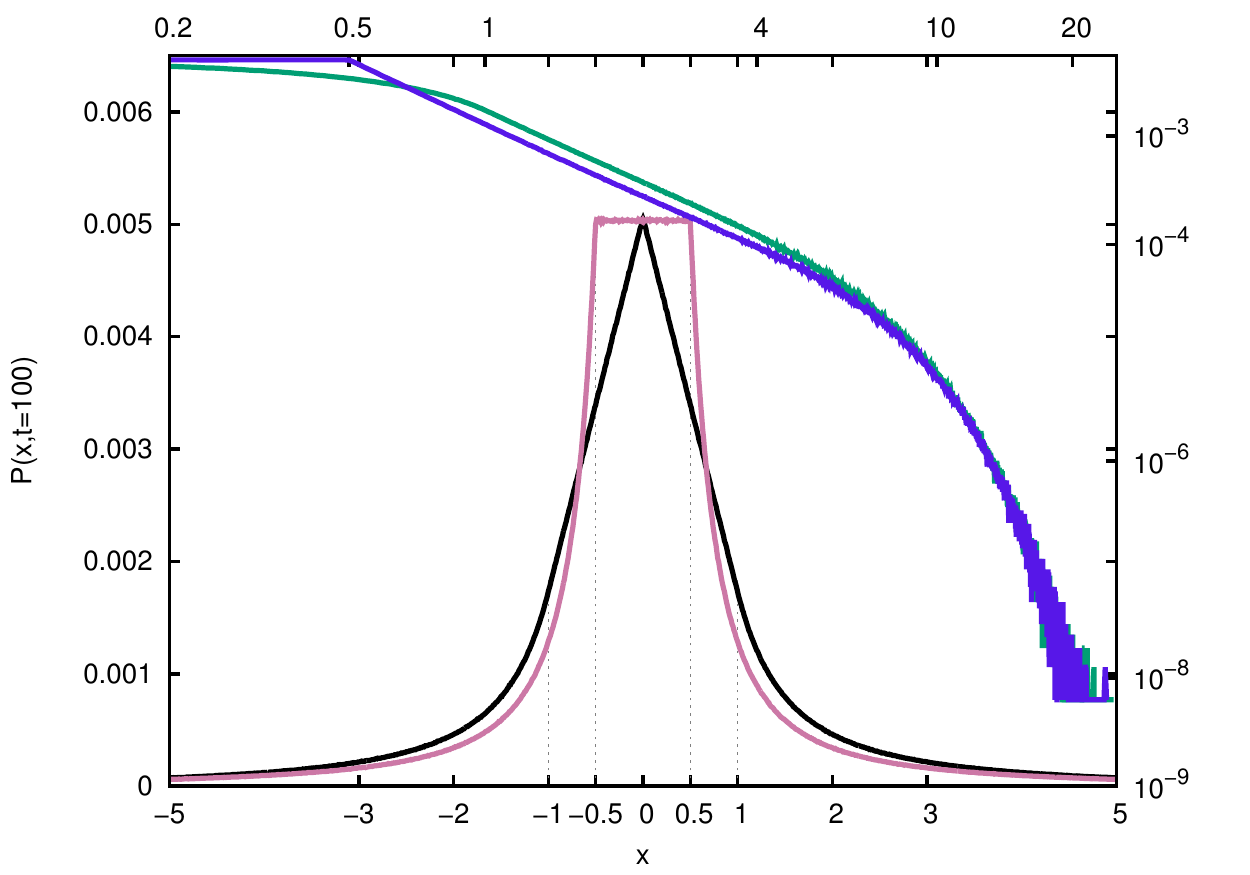}
   \caption{Monte Carlo computation of PDF of displacement
   averaged over segment length $\segl$ for parameters $\alpha=1$, $r_0=1$, $t=100$.
  Two lower curves are linear scale. Two upper curves are double log scale.
  Peaked curves: black linear scale and green log scale are equilibrium density~\eqref{densavposres} 
  averaged over disorder. Flat-top curves: Violet linear scale and blue log scale,
  are average over disorder of~\eqref{solone} with $x_0=0$, ie non-equilibrium $u(x,0) = \delta(x)$.
  Note that averaging over $x_0$ to get the equilibrium distribution increases
  the width of central region by a factor of two.
  }
\label{fig:pdfdisp}
\end{figure}

\myheading{Comparison with other sources of subdiffusion}
Our random confinement model does not manifest weak ergodicity breaking~\cite{Bouchaud1992}, a
characteristic whose presence is known to imply dramatic differences
between EMSD and TEMSD observed in SPT
experiments~\cite{HeBMB2008,LubelskiSK2008,Sokolov2008,Jeon2011,Khoury2011,Kusumi2012}. On
the contrary, the model is in equilibrium at $t=0$, and the increments
of $x(t)$ are stationary so that the TEMSD and EMSD are equal at all
times. Thus, the subdiffusion is due to correlated increments, as is
the case for both DOP and FBM. Our simulations show that the PDF of
the displacement for large times is non-Gaussian as shown in
Fig.~\ref{fig:pdfdisp}, a feature shared by DOP, and CTRW, but not
FBM. RSFC and DOP (but not FBM and CTRW) are examples of diffusion on
disordered, or more broadly, heterogeneous media. 
A model of this type that is closely related to RSFC considers domains of random scale-free size and
diffusivity~\cite{Massignan2014}. Despite this similarity, the latter
model has deeper similarities to other models of heterogeneous
transport coefficients~\cite{Dentz2005,Dentz2010,Kang2011,Kuehn2011,Manzo2015,Cherstvy2013a,Cherstvy2014,Metzler2014}.
When these models do show subdiffusion, it arises from non-stationary
increments, and they are thus more closely related to CTRW.

\myheading{RSFC and percolation} It is interesting to compare RSFC
more closely to DOP. We distinguish two cases: The first is diffusion
of a particle starting at a randomly chosen site on a percolation
process at the critical threshold~\cite{Stauffer1991} (DOP~I).  In the
second case the initial site is a random site on the critical infinite
cluster (DOP~II).  Both DOP~I and DOP~II show subdiffusive EMSD, but
with differing values of the anomalous diffusion exponent $\beta$ in
\eqref{mainmsd}~\cite{BenAvraham1982,Gefen1983,Stauffer1991}.  For
DOP~II, subdiffusion is due to the fractal properties of the infinite
cluster.  However, because the volume fraction of the critical
infinite cluster is zero, it gives no weight in DOP~I where the
initial position may be any point on the lattice. As is the case with
the pure random confinement of RSFC, in DOP~I \textit{every particle
  is confined to a region of finite size}.  Thus, for both RSFC and
DOP~I, the time averaged mean squared displacement (TMSD) ---the time
average of a single trajectory--- tends to a constant at long
times. In DOP~I, the size of the confinement regions (the finite
clusters) has a heavy-tailed, power-law distribution. Thus, as for
RSFC, averaging over an ensemble of uniformly distributed particles
gives an unbounded, subdiffusive MSD, with an additional contribution
from the fractal structure of large clusters.  In other words,
\textit{The fact that $\beta$ differs between (I) the walk on all
  clusters and (II) the walk on the infinite cluster, is due to random,
  scale-free confinement and RSFC represents an abstraction of this
  phenomenon.}

We address a potential issue in simulations and experiment. In
CTRW, a finite number of trajectories show a subdiffusive EMSD no
matter how long their duration, though this implies an ever
increasing spatial domain.  In DOP~I and RSFC, the average over a finite number of trajectories will tend to a
constant EMSD at long times, even for $\alpha< 2$. But, simulations
and SPT trajectories are typically limited to times shorter than the
time $T$ required to explore the entire experimental
domain. Since a heavy-tail ($\alpha<2$) implies that 
there are compartments whose area is of the order of the
experimental domain, the time required to cross these domains will be
of the same order as $T$, and thus power-law subdiffusion will be observed as
shown in Fig.~\ref{fig:msdcurves}.
Note that: In CTRW, no particle is trapped for an infinite time, yet the step
rate decreases toward zero. In RSFC, the motion of \textit{every} particle is
bounded for an infinite time, yet the EMSD grows without bound.

\myheading{Calculation of EMSD}
We denote by $\densone{x_0+y,t}{x_0}$ the solution to \eqref{diffeqn}
with $\segl=1$ and
initial condition $u(x,0) = \delta(x-x_0)$. We translate the origin to $x_0$, and average
over $x_0$ to get the density for displacement from initial position
for a trajectory sampled from the equilibrium distribution,
which we denote by $\densavpos{y,t}$.
For $y\ge 0$ this average is
\begin{equation*}
\overline{u}_{>}(y,t) =  \int_{-1/2}^{1/2-y}  \densone{x_0+y,t}{x_0} \d{x_0}  \quad \text{ for } 0 \le y \le 1.
\end{equation*}
Because $\overline{u}(y,t)$ is symmetric in $y$, we have
\begin{equation}
\label{densavpos}
\densavpos{y,t}  =  \int_{-1/2}^{1/2-\ay} \densone{\ay+x_0,t}{x_0} \d{x_0}  \quad \text{ for } -1 \le y \le 1,
\end{equation}
We denote by $\ensavp{x^2}{\segl}$ the EMSD averaged over $x_0$ for a segment of
length $\segl$. For the segment of unit length, this is given by
\begin{equation}
\begin{aligned}
\label{ensavpdef}
\ensavp{x^2}{1} & = 2 \int_0^1 \densavpos{y,t} y^2 \d{y}.
\end{aligned}
\end{equation}
The EMSD averaged over
the disorder is given by
\begin{equation*}
 \ensavpr{x^2} = \int_{0}^{\infty} \ensavp{x^2}{\segl} P(r) \d{r}.
\end{equation*}
Separating time and space variables, the solution to \eqref{diffeqn} may be written as
an eigenfunction expansion
\begin{gather*}
 u(x;t) = \sum_{n=0}^\infty  \Big[ a_n e^{-4\pi^2 n^2 t} \cos(2\pi n x) \\
  + b_n e^{-4\pi^2(n+1/2)^2 t} \sin(\tpnh) \Big],
\end{gather*}
with coefficients determined by the initial condition and orthogonality of the eigenfunctions.
Using the initial condition $u(x,0) = \delta(x-x_0)$, we find
\begin{equation}
 \label{solone}
  \begin{split}
u(x;x_0,t) 
 =  1 + 2\sum_{n=1}^\infty  e^{-4\pi^2 n^2 t} \cos(2\pi n x)\cos(2\pi n x_0) \\
  +   2\sum_{n=0}^\infty e^{-4\pi^2(n+1/2)^2 t} \sin(\tpnh x) \sin(\tpnh x_0)
  \end{split}
\end{equation}
Evaluating the integrals obtained by substituting \eqref{solone} into \eqref{densavpos} is straightforward and
results in
\begin{equation}
\label{densavposres}
  \begin{aligned}
 \densavpos{y,t} = &  (1-\ay) \\
  + & \sum_{n=1}^\infty  e^{-\pi^2 n^2 t} \left[(1-\ay)\cos(\pi n \ay) - \frac{\sin(n\pi \ay)}{\pi n} \right].
  \end{aligned}
\end{equation}
The asymptotic shape of $\densavpos{y,t}$ is reflected in the central portion of the disorder-averaged
density as seen in Fig.~\ref{fig:pdfdisp}.
Inserting \eqref{densavposres} into \eqref{ensavpdef} we find
\begin{equation*}
\begin{aligned}
\ensavp{x^2}{1} 
   & = \frac{1}{6} - \frac{1}{\pi^4} \sum_{n=0}^\infty  e^{-4\pi^2 (n+1/2)^2 t}  \, \frac{1}{(n+1/2)^4.}
\end{aligned}
\end{equation*}
Setting $t=0$ shows that
\begin{equation}
 \label{zetaone}
 \frac{1}{\pi^4} \sum_{n=0}^\infty \frac{1}{(n+1/2)^4}
 = \frac{1}{6}.
\end{equation}
We now consider a segment of length $r$, rather than $1$, and use \eqref{zetaone} to write
\begin{equation*}
\ensavp{x^2}{\segl} = \frac{r^2}{\pi^4} \sum_{n=0}^\infty \left(1 - e^{-4\pi^2 (n+1/2)^2 t/r^2} \right) \, \frac{1}{(n+1/2)^4},
\end{equation*}
where we have absorbed the initial term into the sum in order to cancel two diverging
quantities below. Averaging over the disorder we have
\begin{equation}
 \label{avdis}
 \begin{aligned}
  \ensavpr{x^2} = & 
   \frac{1}{\pi^4} \sum_{n=0}^{\infty} \frac{1}{(n+1/2)^4} \times \\
  & \int_0^\infty  \left(1 - e^{-4\pi^2 (n+1/2)^2 t/r^2} \right) r^2 P(r) \d{r} \\
 = & 
 \frac{4}{\pi}t^{3/2} \sum_{n=0}^{\infty} \frac{1}{(n+1/2)} \times \\ 
 & \int_0^\infty 
  \left(1 - e^{-y} \right) y^{-5/2} P\left(2\pi(n+1/2)\sqrt{t/y}\right) \d{y}.
 \end{aligned}
\end{equation}
Inserting the Pareto distribution \eqref{pareto} into \eqref{avdis}, we have
\begin{equation}
 \label{genone}
 \begin{aligned}
  \ensavpr{x^2} & =  \\
  & \frac{\alpha r_0^2}{2\pi^4}  \sum_{n=0}^\infty \frac{1}{(n+1/2)^4}
      z^{\frac{2-\alpha}{2}} \int_0^z \left(1-e^{-y}\right) y^{\frac{\alpha-2}{2} - 1} \d{y}\\
  = & \frac{\alpha r_0^2}{2\pi^4}  \sum_{n=0}^\infty \frac{1}{(n+1/2)^4}
      \einbig{\frac{\alpha-2}{2}, \frac{4\pi^2(n+1/2)^2t}{r_0^2}},
 \end{aligned}
\end{equation}
where $z= 4\pi^2(n+1/2)^2t/r_0^2$ and we have defined a generalized entire exponential integral by
\begin{equation}
\label{eindef}
\ein{s,z} = z^{-s} \int_0^z\left(1-e^{-y}\right)y^{s-1} \d{y}.
\end{equation}
The series obtained by expanding the exponential\footnote{%
 $\ein{s,z} = \sum_{k=1}^\infty (-1)^{k+1} \frac{z^k}{k!(k+s)}$
}
is easily seen to be analytic for $s,z\in\mathbb{C}$ except for simple poles
at $s=-1,-2,\ldots$. In particular, \eqref{eindef} is analytic at
$s=0$, where $\ein{0,z} = \ein{z}$, the usual entire exponential integral.
This corresponds to the critical value of the power-law exponent $\alpha=2$.
It can be shown that for $\alpha\ne 2$, $\ein{s,z} = 1/s -z^{-s}\gamma(s,z)$,
where $\gamma(s,z)$ is the lower incomplete gamma function. Large $t$ corresponds
to large $z$, so that, using $\lim_{z\to\infty}\gamma(s,z) = \Gamma(s)$, we find
the asymptotic form
\begin{equation}
 \label{onedsolone}
 \begin{aligned}
  & \ensavpr{x^2}  =
   \frac{\alpha}{\alpha-2}\frac{r_0^2}{6} \\
     & -  \alpha \pi^{-\alpha-2}2^{1-\alpha}r_0^{\alpha} \Gamma\left(\frac{\alpha-2}{2}\right)\zeta(2+\alpha,1/2) t^{\frac{2-\alpha}{2}}
 \end{aligned},
\end{equation}
where $\zeta(s,z)$ is the Hurwitz Zeta function.
Eq.~\eqref{onedsolone} shows that for $\alpha>2$, $\ensavpr{x^2}$ converges to a constant at
long times, while for $1<\alpha<2$, $\ensavpr{x^2}$ shows subdiffusion.
For $\alpha=2$, we use
$\ein{0,z} = \ein{z} = \ln(z) + \gamma + \Gamma(0,z)$ and
$\lim_{z\to\infty}\Gamma(s,z) = 0$  to find the asymptotic solution
\begin{equation}
 \label{logsol}
 \begin{aligned}
  \ensavpr{x^2} & = 
     \frac{r_0^2}{6}
     \left[\gamma +  2\ln\left(\frac{2\pi}{r_0}\right)
  - \frac{12\partial_s\zeta(4,1/2)}{\pi^4}  + \ln(t) \right]
  \end{aligned}
\end{equation}
where $\partial_s\zeta(s,1/2)|_{s=4} \approx 10.9697$.
Equations \eqref{onedsolone} and \eqref{logsol} are the goals of the calculations and are
shown in Fig.~\ref{fig:msdcurves}.

\myheading{Non-equilibrium initial conditions}
Although we assumed equilibrium above, subdiffusion is
also observed for other initial distributions.
We performed calculations analogous to those above
for the initial distribution $u(x,0) = \delta(x)$, that is, each particle begins at the center
of the confinement domain, and obtained the asymptotic form for $\alpha \ne 2$
\begin{equation}
 \label{onedasympgen}
\ensavpr{x^2} =  \frac{\alpha}{\alpha-2}\frac{r_0^2}{12}
     - \alpha \pi^{-\alpha}2^{1-\alpha}r_0^{\alpha} \Gamma\left(\frac{\alpha-2}{2}\right)\eta(\alpha) t^{\frac{2-\alpha}{2}},
\end{equation}
where $\eta(x)$ is the Dirichlet eta function.
For $\alpha = 2$ we found
\begin{equation*}
 \begin{aligned}
 \ensavpr{x^2}   =  
    \frac{r_0^2}{6} \bigg\{ \frac{\ln(t)}{2}  -\frac{6\zeta'(2)}{\pi^2} + \ln(\pi)
   + \frac{\gamma}{2} -\ln(r_0) \bigg\},
 \end{aligned}
\end{equation*}
where $\zeta(x)$ is the Riemann zeta function.
In two dimensions, ie diffusion on disks of random radius, we obtained the asymptotic forms, for $\alpha\ne 2$
\begin{equation*}
\ensavpr{\rho^2} =
 \frac{\alpha}{\alpha-2}\frac{r_0^2}{2} - 2\alpha r_0^{\alpha} 
  \Gamma\left(\frac{\alpha-2}{2}\right) \njsum{\alpha}
  t^{-\frac{\alpha-2}{2}},
\end{equation*}
where
$\zJ{\nu,n}$ is the $n$th positive zero of the Bessel function of the first kind $J_\nu$,
and $\nsumJnu{\beta}$ is
\begin{equation*}
 \nsumJnu{\beta} = \bsumbess{\nu}{\nu+1}{\beta-\nu+1},
\end{equation*}
and $\rho(t)$ is the radial density, ie averaged over azimuthal angle.
For $\alpha=2$, we found
\begin{equation*}
\begin{aligned}
\ensavpr{\rho^2} & =  \frac{1}{2}r_0^2\left\{ \gamma -\ln(r_0^2) -16 \partial_{\beta}\nsumJ{1}{2} +\ln(t) \right\},
\end{aligned}
\end{equation*}
where $\partial_{\beta}\nsumJ{1}{2} \approx 0.147342$.

\myheading{Conclusion} We have demonstrated that heavy-tailed random
scale-free confinement gives rise to a subdiffusive EMSD. We have discussed its relation to other
sources of anomalous diffusion, in particular its presence in certain diffusive processes on percolation.
A number of questions remain, including: Are there random potential fields that
lead to RSFC ?  What is nature of the propagator (PDF) of RSFC ? \ldots or of autocorrelation of the displacement?
What is the anomalous exponent when RSFC is combined with other sources of anomalous
diffusion ?

\vspace{10pt}
\begin{acknowledgements}
  \myheading{Acknowledgments}
   This work was supported by the European Research Council (ERC)
  through the project MHetScale (Contract number 617511),
  Fundaci\'o Cellex, 
 ERC AdG Osyris, and
  Spanish Ministry Project FOQUS (FIS2013-46768-P)
\end{acknowledgements}

\bibliography{randbarrier}

\begin{thebibliography}{50}%
\makeatletter
\providecommand \@ifxundefined [1]{%
 \@ifx{#1\undefined}
}%
\providecommand \@ifnum [1]{%
 \ifnum #1\expandafter \@firstoftwo
 \else \expandafter \@secondoftwo
 \fi
}%
\providecommand \@ifx [1]{%
 \ifx #1\expandafter \@firstoftwo
 \else \expandafter \@secondoftwo
 \fi
}%
\providecommand \natexlab [1]{#1}%
\providecommand \enquote  [1]{``#1''}%
\providecommand \bibnamefont  [1]{#1}%
\providecommand \bibfnamefont [1]{#1}%
\providecommand \citenamefont [1]{#1}%
\providecommand \href@noop [0]{\@secondoftwo}%
\providecommand \href [0]{\begingroup \@sanitize@url \@href}%
\providecommand \@href[1]{\@@startlink{#1}\@@href}%
\providecommand \@@href[1]{\endgroup#1\@@endlink}%
\providecommand \@sanitize@url [0]{\catcode `\\12\catcode `\$12\catcode
  `\&12\catcode `\#12\catcode `\^12\catcode `\_12\catcode `\%12\relax}%
\providecommand \@@startlink[1]{}%
\providecommand \@@endlink[0]{}%
\providecommand \url  [0]{\begingroup\@sanitize@url \@url }%
\providecommand \@url [1]{\endgroup\@href {#1}{\urlprefix }}%
\providecommand \urlprefix  [0]{URL }%
\providecommand \Eprint [0]{\href }%
\providecommand \doibase [0]{http://dx.doi.org/}%
\providecommand \selectlanguage [0]{\@gobble}%
\providecommand \bibinfo  [0]{\@secondoftwo}%
\providecommand \bibfield  [0]{\@secondoftwo}%
\providecommand \translation [1]{[#1]}%
\providecommand \BibitemOpen [0]{}%
\providecommand \bibitemStop [0]{}%
\providecommand \bibitemNoStop [0]{.\EOS\space}%
\providecommand \EOS [0]{\spacefactor3000\relax}%
\providecommand \BibitemShut  [1]{\csname bibitem#1\endcsname}%
\let\auto@bib@innerbib\@empty
\bibitem [{\citenamefont {Havlin}\ and\ \citenamefont
  {Ben-Avraham}(1987)}]{Havlin1987}%
  \BibitemOpen
  \bibfield  {author} {\bibinfo {author} {\bibfnamefont {S.}~\bibnamefont
  {Havlin}}\ and\ \bibinfo {author} {\bibfnamefont {D.}~\bibnamefont
  {Ben-Avraham}},\ }\href {\doibase 10.1080/00018738700101072} {\bibfield
  {journal} {\bibinfo  {journal} {Adv. Phys.}\ }\textbf {\bibinfo {volume}
  {36}},\ \bibinfo {pages} {695} (\bibinfo {year} {1987})}\BibitemShut
  {NoStop}%
\bibitem [{\citenamefont {{Bouchaud}}\ and\ \citenamefont
  {{Georges}}(1990)}]{Bouchaud1990}%
  \BibitemOpen
  \bibfield  {author} {\bibinfo {author} {\bibfnamefont {J.-P.}\ \bibnamefont
  {{Bouchaud}}}\ and\ \bibinfo {author} {\bibfnamefont {A.}~\bibnamefont
  {{Georges}}},\ }\href {\doibase 10.1016/0370-1573(90)90099-N} {\bibfield
  {journal} {\bibinfo  {journal} {Phys. Rep.}\ }\textbf {\bibinfo {volume}
  {195}},\ \bibinfo {pages} {127} (\bibinfo {year} {1990})}\BibitemShut
  {NoStop}%
\bibitem [{\citenamefont {Metzler}\ and\ \citenamefont
  {Klafter}(2004)}]{Metzler2004}%
  \BibitemOpen
  \bibfield  {author} {\bibinfo {author} {\bibfnamefont {R.}~\bibnamefont
  {Metzler}}\ and\ \bibinfo {author} {\bibfnamefont {J.}~\bibnamefont
  {Klafter}},\ }\href {http://stacks.iop.org/0305-4470/37/i=31/a=R01}
  {\bibfield  {journal} {\bibinfo  {journal} {J. Phys. A}\ }\textbf {\bibinfo
  {volume} {37}},\ \bibinfo {pages} {R161} (\bibinfo {year}
  {2004})}\BibitemShut {NoStop}%
\bibitem [{\citenamefont {Barkai}\ \emph {et~al.}(2012)\citenamefont {Barkai},
  \citenamefont {Garini},\ and\ \citenamefont {Metzler}}]{Barkai2012a}%
  \BibitemOpen
  \bibfield  {author} {\bibinfo {author} {\bibfnamefont {E.}~\bibnamefont
  {Barkai}}, \bibinfo {author} {\bibfnamefont {Y.}~\bibnamefont {Garini}}, \
  and\ \bibinfo {author} {\bibfnamefont {R.}~\bibnamefont {Metzler}},\ }\href
  {\doibase 10.1063/PT.3.1677} {\bibfield  {journal} {\bibinfo  {journal}
  {Phys. Today}\ }\textbf {\bibinfo {volume} {65}},\ \bibinfo {pages} {29}
  (\bibinfo {year} {2012})}\BibitemShut {NoStop}%
\bibitem [{\citenamefont {H\"{o}fling}\ and\ \citenamefont
  {Franosch}(2013)}]{Hoefling2013}%
  \BibitemOpen
  \bibfield  {author} {\bibinfo {author} {\bibfnamefont {F.}~\bibnamefont
  {H\"{o}fling}}\ and\ \bibinfo {author} {\bibfnamefont {T.}~\bibnamefont
  {Franosch}},\ }\href {\doibase 10.1088/0034-4885/76/4/046602} {\bibfield
  {journal} {\bibinfo  {journal} {Rep. Prog. Phys.}\ }\textbf {\bibinfo
  {volume} {76}},\ \bibinfo {pages} {046602} (\bibinfo {year}
  {2013})}\BibitemShut {NoStop}%
\bibitem [{\citenamefont {Metzler}\ \emph {et~al.}(2014)\citenamefont
  {Metzler}, \citenamefont {Jeon}, \citenamefont {Cherstvy},\ and\
  \citenamefont {Barkai}}]{Metzler2014}%
  \BibitemOpen
  \bibfield  {author} {\bibinfo {author} {\bibfnamefont {R.}~\bibnamefont
  {Metzler}}, \bibinfo {author} {\bibfnamefont {J.-H.}\ \bibnamefont {Jeon}},
  \bibinfo {author} {\bibfnamefont {A.~G.}\ \bibnamefont {Cherstvy}}, \ and\
  \bibinfo {author} {\bibfnamefont {E.}~\bibnamefont {Barkai}},\ }\href
  {\doibase 10.1039/C4CP03465A} {\bibfield  {journal} {\bibinfo  {journal}
  {Phys. Chem. Chem. Phys.}\ }\textbf {\bibinfo {volume} {16}},\ \bibinfo
  {pages} {24128} (\bibinfo {year} {2014})}\BibitemShut {NoStop}%
\bibitem [{\citenamefont {Scher}\ and\ \citenamefont
  {Montroll}(1975)}]{ScherMontroll75}%
  \BibitemOpen
  \bibfield  {author} {\bibinfo {author} {\bibfnamefont {H.}~\bibnamefont
  {Scher}}\ and\ \bibinfo {author} {\bibfnamefont {E.~W.}\ \bibnamefont
  {Montroll}},\ }\href {\doibase 10.1103/PhysRevB.12.2455} {\bibfield
  {journal} {\bibinfo  {journal} {Phys. Rev. B}\ }\textbf {\bibinfo {volume}
  {12}},\ \bibinfo {pages} {2455} (\bibinfo {year} {1975})}\BibitemShut
  {NoStop}%
\bibitem [{\citenamefont {Klafter}\ \emph {et~al.}(1987)\citenamefont
  {Klafter}, \citenamefont {Blumen},\ and\ \citenamefont
  {Shlesinger}}]{Klafter1987}%
  \BibitemOpen
  \bibfield  {author} {\bibinfo {author} {\bibfnamefont {J.}~\bibnamefont
  {Klafter}}, \bibinfo {author} {\bibfnamefont {A.}~\bibnamefont {Blumen}}, \
  and\ \bibinfo {author} {\bibfnamefont {M.~F.}\ \bibnamefont {Shlesinger}},\
  }\href {\doibase 10.1103/PhysRevA.35.3081} {\bibfield  {journal} {\bibinfo
  {journal} {Phys. Rev. A}\ }\textbf {\bibinfo {volume} {35}},\ \bibinfo
  {pages} {3081} (\bibinfo {year} {1987})}\BibitemShut {NoStop}%
\bibitem [{\citenamefont {Klafter}\ and\ \citenamefont
  {Sokolov}(2011)}]{Klafter2011}%
  \BibitemOpen
  \bibfield  {author} {\bibinfo {author} {\bibfnamefont {J.}~\bibnamefont
  {Klafter}}\ and\ \bibinfo {author} {\bibfnamefont {I.~M.}\ \bibnamefont
  {Sokolov}},\ }\href@noop {} {\emph {\bibinfo {title} {First Steps in Random
  Walks}}}\ (\bibinfo  {publisher} {Oxford University Press},\ \bibinfo
  {address} {Oxford},\ \bibinfo {year} {2011})\BibitemShut {NoStop}%
\bibitem [{\citenamefont {Kolmogorov}(1940)}]{Kolmogorov1940}%
  \BibitemOpen
  \bibfield  {author} {\bibinfo {author} {\bibfnamefont {A.~N.}\ \bibnamefont
  {Kolmogorov}},\ }\href@noop {} {\bibfield  {journal} {\bibinfo  {journal}
  {Dokl. Akad. Nauk SSSR}\ }\textbf {\bibinfo {volume} {26}} (\bibinfo {year}
  {1940})}\BibitemShut {NoStop}%
\bibitem [{\citenamefont {Mandelbrot}\ and\ \citenamefont
  {Ness}(1968)}]{Mandelbrot1968}%
  \BibitemOpen
  \bibfield  {author} {\bibinfo {author} {\bibfnamefont {B.~B.}\ \bibnamefont
  {Mandelbrot}}\ and\ \bibinfo {author} {\bibfnamefont {J.~W.~V.}\ \bibnamefont
  {Ness}},\ }\href {\doibase 10.1137/1010093} {\bibfield  {journal} {\bibinfo
  {journal} {SIAM Rev.}\ }\textbf {\bibinfo {volume} {10}},\ \bibinfo {pages}
  {422} (\bibinfo {year} {1968})}\BibitemShut {NoStop}%
\bibitem [{\citenamefont {Stauffer}\ and\ \citenamefont
  {Aharony}(1991)}]{Stauffer1991}%
  \BibitemOpen
  \bibfield  {author} {\bibinfo {author} {\bibfnamefont {D.}~\bibnamefont
  {Stauffer}}\ and\ \bibinfo {author} {\bibfnamefont {A.}~\bibnamefont
  {Aharony}},\ }\href@noop {} {\emph {\bibinfo {title} {Introduction to
  Percolation Theory}}}\ (\bibinfo  {publisher} {Taylor and Francis},\ \bibinfo
  {address} {London},\ \bibinfo {year} {1991})\BibitemShut {NoStop}%
\bibitem [{\citenamefont {Meroz}\ \emph {et~al.}(2010)\citenamefont {Meroz},
  \citenamefont {Sokolov},\ and\ \citenamefont {Klafter}}]{Meroz2010}%
  \BibitemOpen
  \bibfield  {author} {\bibinfo {author} {\bibfnamefont {Y.}~\bibnamefont
  {Meroz}}, \bibinfo {author} {\bibfnamefont {I.~M.}\ \bibnamefont {Sokolov}},
  \ and\ \bibinfo {author} {\bibfnamefont {J.}~\bibnamefont {Klafter}},\ }\href
  {\doibase 10.1103/PhysRevE.81.010101} {\bibfield  {journal} {\bibinfo
  {journal} {Phys. Rev. E}\ }\textbf {\bibinfo {volume} {81}},\ \bibinfo
  {pages} {010101} (\bibinfo {year} {2010})}\BibitemShut {NoStop}%
\bibitem [{\citenamefont {Weigel}\ \emph {et~al.}(2011)\citenamefont {Weigel},
  \citenamefont {Simon}, \citenamefont {Tamkun},\ and\ \citenamefont
  {Krapf}}]{Weigel2011}%
  \BibitemOpen
  \bibfield  {author} {\bibinfo {author} {\bibfnamefont {A.~V.}\ \bibnamefont
  {Weigel}}, \bibinfo {author} {\bibfnamefont {B.}~\bibnamefont {Simon}},
  \bibinfo {author} {\bibfnamefont {M.~M.}\ \bibnamefont {Tamkun}}, \ and\
  \bibinfo {author} {\bibfnamefont {D.}~\bibnamefont {Krapf}},\ }\href
  {\doibase 10.1073/pnas.1016325108} {\bibfield  {journal} {\bibinfo  {journal}
  {Proc. Natl. Acad. Sci. USA}\ }\textbf {\bibinfo {volume} {108}},\ \bibinfo
  {pages} {6438} (\bibinfo {year} {2011})}\BibitemShut {NoStop}%
\bibitem [{\citenamefont {Magdziarz}\ \emph {et~al.}(2009)\citenamefont
  {Magdziarz}, \citenamefont {Weron}, \citenamefont {Burnecki},\ and\
  \citenamefont {Klafter}}]{Magdziarz2009}%
  \BibitemOpen
  \bibfield  {author} {\bibinfo {author} {\bibfnamefont {M.}~\bibnamefont
  {Magdziarz}}, \bibinfo {author} {\bibfnamefont {A.}~\bibnamefont {Weron}},
  \bibinfo {author} {\bibfnamefont {K.}~\bibnamefont {Burnecki}}, \ and\
  \bibinfo {author} {\bibfnamefont {J.}~\bibnamefont {Klafter}},\ }\href
  {\doibase 10.1103/PhysRevLett.103.180602} {\bibfield  {journal} {\bibinfo
  {journal} {Phys. Rev. Lett.}\ }\textbf {\bibinfo {volume} {103}},\ \bibinfo
  {pages} {180602} (\bibinfo {year} {2009})}\BibitemShut {NoStop}%
\bibitem [{\citenamefont {Magdziarz}\ and\ \citenamefont
  {Weron}(2011)}]{Magdziarz2011}%
  \BibitemOpen
  \bibfield  {author} {\bibinfo {author} {\bibfnamefont {M.}~\bibnamefont
  {Magdziarz}}\ and\ \bibinfo {author} {\bibfnamefont {A.}~\bibnamefont
  {Weron}},\ }\href {\doibase 10.1103/PhysRevE.84.051138} {\bibfield  {journal}
  {\bibinfo  {journal} {Phys. Rev. E}\ }\textbf {\bibinfo {volume} {84}},\
  \bibinfo {pages} {051138} (\bibinfo {year} {2011})}\BibitemShut {NoStop}%
\bibitem [{\citenamefont {Jeon}\ \emph {et~al.}(2013)\citenamefont {Jeon},
  \citenamefont {Barkai},\ and\ \citenamefont {Metzler}}]{Jeon2013}%
  \BibitemOpen
  \bibfield  {author} {\bibinfo {author} {\bibfnamefont {J.-H.}\ \bibnamefont
  {Jeon}}, \bibinfo {author} {\bibfnamefont {E.}~\bibnamefont {Barkai}}, \ and\
  \bibinfo {author} {\bibfnamefont {R.}~\bibnamefont {Metzler}},\ }\href
  {\doibase http://dx.doi.org/10.1063/1.4816635} {\bibfield  {journal}
  {\bibinfo  {journal} {The Journal of Chemical Physics}\ }\textbf {\bibinfo
  {volume} {139}},\ \bibinfo {eid} {121916} (\bibinfo {year}
  {2013})}\BibitemShut {NoStop}%
\bibitem [{\citenamefont {Meroz}\ \emph {et~al.}(2013)\citenamefont {Meroz},
  \citenamefont {Sokolov},\ and\ \citenamefont {Klafter}}]{Meroz2013}%
  \BibitemOpen
  \bibfield  {author} {\bibinfo {author} {\bibfnamefont {Y.}~\bibnamefont
  {Meroz}}, \bibinfo {author} {\bibfnamefont {I.~M.}\ \bibnamefont {Sokolov}},
  \ and\ \bibinfo {author} {\bibfnamefont {J.}~\bibnamefont {Klafter}},\ }\href
  {\doibase 10.1103/PhysRevLett.110.090601} {\bibfield  {journal} {\bibinfo
  {journal} {Phys. Rev. Lett.}\ }\textbf {\bibinfo {volume} {110}},\ \bibinfo
  {pages} {090601} (\bibinfo {year} {2013})}\BibitemShut {NoStop}%
\bibitem [{\citenamefont {Meilhac}\ \emph {et~al.}(2006)\citenamefont
  {Meilhac}, \citenamefont {Le~Guyader}, \citenamefont {Salom\'e},\ and\
  \citenamefont {Destainville}}]{Meilhac2006}%
  \BibitemOpen
  \bibfield  {author} {\bibinfo {author} {\bibfnamefont {N.}~\bibnamefont
  {Meilhac}}, \bibinfo {author} {\bibfnamefont {L.}~\bibnamefont {Le~Guyader}},
  \bibinfo {author} {\bibfnamefont {L.}~\bibnamefont {Salom\'e}}, \ and\
  \bibinfo {author} {\bibfnamefont {N.}~\bibnamefont {Destainville}},\ }\href
  {\doibase 10.1103/PhysRevE.73.011915} {\bibfield  {journal} {\bibinfo
  {journal} {Phys. Rev. E}\ }\textbf {\bibinfo {volume} {73}},\ \bibinfo
  {pages} {011915} (\bibinfo {year} {2006})}\BibitemShut {NoStop}%
\bibitem [{\citenamefont {Condamin}\ \emph {et~al.}(2007)\citenamefont
  {Condamin}, \citenamefont {B\'{e}nichou},\ and\ \citenamefont
  {Klafter}}]{Condamin2007}%
  \BibitemOpen
  \bibfield  {author} {\bibinfo {author} {\bibfnamefont {S.}~\bibnamefont
  {Condamin}}, \bibinfo {author} {\bibfnamefont {O.}~\bibnamefont
  {B\'{e}nichou}}, \ and\ \bibinfo {author} {\bibfnamefont {J.}~\bibnamefont
  {Klafter}},\ }\href {\doibase 10.1103/PhysRevLett.98.250602} {\bibfield
  {journal} {\bibinfo  {journal} {Phys. Rev. Lett.}\ }\textbf {\bibinfo
  {volume} {98}},\ \bibinfo {pages} {250602} (\bibinfo {year}
  {2007})}\BibitemShut {NoStop}%
\bibitem [{\citenamefont {Destainville}\ \emph {et~al.}(2008)\citenamefont
  {Destainville}, \citenamefont {Sauli\'{e}re},\ and\ \citenamefont
  {Salom\'{e}}}]{Destainville2008}%
  \BibitemOpen
  \bibfield  {author} {\bibinfo {author} {\bibfnamefont {N.}~\bibnamefont
  {Destainville}}, \bibinfo {author} {\bibfnamefont {A.}~\bibnamefont
  {Sauli\'{e}re}}, \ and\ \bibinfo {author} {\bibfnamefont {L.}~\bibnamefont
  {Salom\'{e}}},\ }\href {\doibase 10.1529/biophysj.108.136739} {\bibfield
  {journal} {\bibinfo  {journal} {Biophys. J.}\ }\textbf {\bibinfo {volume}
  {95}},\ \bibinfo {pages} {3117} (\bibinfo {year} {2008})}\BibitemShut
  {NoStop}%
\bibitem [{\citenamefont {Condamin}\ \emph {et~al.}(2008)\citenamefont
  {Condamin}, \citenamefont {Tejedor}, \citenamefont {Voituriez}, \citenamefont
  {B\'{e}nichou},\ and\ \citenamefont {Klafter}}]{Condamin2008}%
  \BibitemOpen
  \bibfield  {author} {\bibinfo {author} {\bibfnamefont {S.}~\bibnamefont
  {Condamin}}, \bibinfo {author} {\bibfnamefont {V.}~\bibnamefont {Tejedor}},
  \bibinfo {author} {\bibfnamefont {R.}~\bibnamefont {Voituriez}}, \bibinfo
  {author} {\bibfnamefont {O.}~\bibnamefont {B\'{e}nichou}}, \ and\ \bibinfo
  {author} {\bibfnamefont {J.}~\bibnamefont {Klafter}},\ }\href {\doibase
  10.1073/pnas.0712158105} {\bibfield  {journal} {\bibinfo  {journal} {Proc.
  Natl. Acad. Sci. USA}\ }\textbf {\bibinfo {volume} {105}},\ \bibinfo {pages}
  {5675} (\bibinfo {year} {2008})}\BibitemShut {NoStop}%
\bibitem [{\citenamefont {Denisov}\ \emph {et~al.}(2008)\citenamefont
  {Denisov}, \citenamefont {Horsthemke},\ and\ \citenamefont
  {H\"{a}nggi}}]{Denisov2008}%
  \BibitemOpen
  \bibfield  {author} {\bibinfo {author} {\bibfnamefont {S.~I.}\ \bibnamefont
  {Denisov}}, \bibinfo {author} {\bibfnamefont {W.}~\bibnamefont {Horsthemke}},
  \ and\ \bibinfo {author} {\bibfnamefont {P.}~\bibnamefont {H\"{a}nggi}},\
  }\href {\doibase 10.1103/PhysRevE.77.061112} {\bibfield  {journal} {\bibinfo
  {journal} {Physical Review E}\ }\textbf {\bibinfo {volume} {77}},\ \bibinfo
  {pages} {061112} (\bibinfo {year} {2008})}\BibitemShut {NoStop}%
\bibitem [{\citenamefont {Burada}\ \emph {et~al.}(2009)\citenamefont {Burada},
  \citenamefont {H\"{a}nggi}, \citenamefont {Marchesoni}, \citenamefont
  {Schmid},\ and\ \citenamefont {Talkner}}]{Burada2009}%
  \BibitemOpen
  \bibfield  {author} {\bibinfo {author} {\bibfnamefont {P.~S.}\ \bibnamefont
  {Burada}}, \bibinfo {author} {\bibfnamefont {P.}~\bibnamefont {H\"{a}nggi}},
  \bibinfo {author} {\bibfnamefont {F.}~\bibnamefont {Marchesoni}}, \bibinfo
  {author} {\bibfnamefont {G.}~\bibnamefont {Schmid}}, \ and\ \bibinfo {author}
  {\bibfnamefont {P.}~\bibnamefont {Talkner}},\ }\href {\doibase
  10.1002/cphc.200800526} {\bibfield  {journal} {\bibinfo  {journal}
  {ChemPhysChem}\ }\textbf {\bibinfo {volume} {10}},\ \bibinfo {pages} {45}
  (\bibinfo {year} {2009})}\BibitemShut {NoStop}%
\bibitem [{\citenamefont {Neusius}\ \emph {et~al.}(2009)\citenamefont
  {Neusius}, \citenamefont {Sokolov},\ and\ \citenamefont
  {Smith}}]{Neusius2009}%
  \BibitemOpen
  \bibfield  {author} {\bibinfo {author} {\bibfnamefont {T.}~\bibnamefont
  {Neusius}}, \bibinfo {author} {\bibfnamefont {I.~M.}\ \bibnamefont
  {Sokolov}}, \ and\ \bibinfo {author} {\bibfnamefont {J.~C.}\ \bibnamefont
  {Smith}},\ }\href {\doibase 10.1103/PhysRevE.80.011109} {\bibfield  {journal}
  {\bibinfo  {journal} {Phys. Rev. E}\ }\textbf {\bibinfo {volume} {80}},\
  \bibinfo {pages} {011109} (\bibinfo {year} {2009})}\BibitemShut {NoStop}%
\bibitem [{\citenamefont {Magdziarz}\ and\ \citenamefont
  {Klafter}(2010)}]{Magdziarz2010}%
  \BibitemOpen
  \bibfield  {author} {\bibinfo {author} {\bibfnamefont {M.}~\bibnamefont
  {Magdziarz}}\ and\ \bibinfo {author} {\bibfnamefont {J.}~\bibnamefont
  {Klafter}},\ }\href {\doibase 10.1103/PhysRevE.82.011129} {\bibfield
  {journal} {\bibinfo  {journal} {Phys. Rev. E}\ }\textbf {\bibinfo {volume}
  {82}},\ \bibinfo {pages} {011129} (\bibinfo {year} {2010})}\BibitemShut
  {NoStop}%
\bibitem [{\citenamefont {Jeon}\ and\ \citenamefont
  {Metzler}(2012)}]{Jeon2012}%
  \BibitemOpen
  \bibfield  {author} {\bibinfo {author} {\bibfnamefont {J.-H.}\ \bibnamefont
  {Jeon}}\ and\ \bibinfo {author} {\bibfnamefont {R.}~\bibnamefont {Metzler}},\
  }\href {\doibase 10.1103/PhysRevE.85.021147} {\bibfield  {journal} {\bibinfo
  {journal} {Phys. Rev. E}\ }\textbf {\bibinfo {volume} {85}},\ \bibinfo
  {pages} {021147} (\bibinfo {year} {2012})}\BibitemShut {NoStop}%
\bibitem [{\citenamefont {Bruna}\ and\ \citenamefont
  {Chapman}(2014)}]{Bruna2014}%
  \BibitemOpen
  \bibfield  {author} {\bibinfo {author} {\bibfnamefont {M.}~\bibnamefont
  {Bruna}}\ and\ \bibinfo {author} {\bibfnamefont {S.}~\bibnamefont
  {Chapman}},\ }\href {\doibase 10.1007/s11538-013-9847-0} {\bibfield
  {journal} {\bibinfo  {journal} {Bulletin of Mathematical Biology}\ }\textbf
  {\bibinfo {volume} {76}},\ \bibinfo {pages} {947} (\bibinfo {year}
  {2014})}\BibitemShut {NoStop}%
\bibitem [{\citenamefont {Toli\'{c}-N\o{}rrelykke}\ \emph
  {et~al.}(2004)\citenamefont {Toli\'{c}-N\o{}rrelykke}, \citenamefont
  {Munteanu}, \citenamefont {Thon}, \citenamefont {Oddershede},\ and\
  \citenamefont {Berg-S\o{}rensen}}]{Tolic04}%
  \BibitemOpen
  \bibfield  {author} {\bibinfo {author} {\bibfnamefont {I.~M.}\ \bibnamefont
  {Toli\'{c}-N\o{}rrelykke}}, \bibinfo {author} {\bibfnamefont {E.-L.}\
  \bibnamefont {Munteanu}}, \bibinfo {author} {\bibfnamefont {G.}~\bibnamefont
  {Thon}}, \bibinfo {author} {\bibfnamefont {L.}~\bibnamefont {Oddershede}}, \
  and\ \bibinfo {author} {\bibfnamefont {K.}~\bibnamefont {Berg-S\o{}rensen}},\
  }\href {\doibase 10.1103/PhysRevLett.93.078102} {\bibfield  {journal}
  {\bibinfo  {journal} {Phys. Rev. Lett.}\ }\textbf {\bibinfo {volume} {93}},\
  \bibinfo {pages} {078102} (\bibinfo {year} {2004})}\BibitemShut {NoStop}%
\bibitem [{\citenamefont {Golding}\ and\ \citenamefont
  {Cox}(2006)}]{Golding06}%
  \BibitemOpen
  \bibfield  {author} {\bibinfo {author} {\bibfnamefont {I.}~\bibnamefont
  {Golding}}\ and\ \bibinfo {author} {\bibfnamefont {E.~C.}\ \bibnamefont
  {Cox}},\ }\href {\doibase 10.1103/PhysRevLett.96.098102} {\bibfield
  {journal} {\bibinfo  {journal} {Phys. Rev. Lett.}\ }\textbf {\bibinfo
  {volume} {96}},\ \bibinfo {pages} {098102} (\bibinfo {year}
  {2006})}\BibitemShut {NoStop}%
\bibitem [{\citenamefont {Burov}\ \emph {et~al.}(2011)\citenamefont {Burov},
  \citenamefont {Jeon}, \citenamefont {Metzler},\ and\ \citenamefont
  {Barkai}}]{Burov2011}%
  \BibitemOpen
  \bibfield  {author} {\bibinfo {author} {\bibfnamefont {S.}~\bibnamefont
  {Burov}}, \bibinfo {author} {\bibfnamefont {J.-H.}\ \bibnamefont {Jeon}},
  \bibinfo {author} {\bibfnamefont {R.}~\bibnamefont {Metzler}}, \ and\
  \bibinfo {author} {\bibfnamefont {E.}~\bibnamefont {Barkai}},\ }\href
  {\doibase 10.1039/C0CP01879A} {\bibfield  {journal} {\bibinfo  {journal}
  {Phys. Chem. Chem. Phys.}\ }\textbf {\bibinfo {volume} {13}},\ \bibinfo
  {pages} {1800} (\bibinfo {year} {2011})}\BibitemShut {NoStop}%
\bibitem [{\citenamefont {Giannone}\ \emph {et~al.}(2013)\citenamefont
  {Giannone}, \citenamefont {Hosy}, \citenamefont {Sibarita}, \citenamefont
  {Choquet},\ and\ \citenamefont {Cognet}}]{Giannone13}%
  \BibitemOpen
  \bibfield  {author} {\bibinfo {author} {\bibfnamefont {G.}~\bibnamefont
  {Giannone}}, \bibinfo {author} {\bibfnamefont {E.}~\bibnamefont {Hosy}},
  \bibinfo {author} {\bibfnamefont {J.-B.}\ \bibnamefont {Sibarita}}, \bibinfo
  {author} {\bibfnamefont {D.}~\bibnamefont {Choquet}}, \ and\ \bibinfo
  {author} {\bibfnamefont {L.}~\bibnamefont {Cognet}},\ }in\ \href {\doibase
  10.1007/978-1-62703-137-0_7} {\emph {\bibinfo {booktitle} {Nanoimaging}}},\
  \bibinfo {series} {Methods in Molecular Biology}, Vol.\ \bibinfo {volume}
  {950},\ \bibinfo {editor} {edited by\ \bibinfo {editor} {\bibfnamefont
  {A.~A.}\ \bibnamefont {Sousa}}\ and\ \bibinfo {editor} {\bibfnamefont
  {M.~J.}\ \bibnamefont {Kruhlak}}}\ (\bibinfo  {publisher} {Humana Press},\
  \bibinfo {address} {New York},\ \bibinfo {year} {2013})\ pp.\ \bibinfo
  {pages} {95--110}\BibitemShut {NoStop}%
\bibitem [{\citenamefont {{J. P. Bouchaud}}(1992)}]{Bouchaud1992}%
  \BibitemOpen
  \bibfield  {author} {\bibinfo {author} {\bibnamefont {{J. P. Bouchaud}}},\
  }\href {\doibase 10.1051/jp1:1992238} {\bibfield  {journal} {\bibinfo
  {journal} {J. Phys. I France}\ }\textbf {\bibinfo {volume} {2}},\ \bibinfo
  {pages} {1705} (\bibinfo {year} {1992})}\BibitemShut {NoStop}%
\bibitem [{\citenamefont {He}\ \emph {et~al.}(2008)\citenamefont {He},
  \citenamefont {Burov}, \citenamefont {Metzler},\ and\ \citenamefont
  {Barkai}}]{HeBMB2008}%
  \BibitemOpen
  \bibfield  {author} {\bibinfo {author} {\bibfnamefont {Y.}~\bibnamefont
  {He}}, \bibinfo {author} {\bibfnamefont {S.}~\bibnamefont {Burov}}, \bibinfo
  {author} {\bibfnamefont {R.}~\bibnamefont {Metzler}}, \ and\ \bibinfo
  {author} {\bibfnamefont {E.}~\bibnamefont {Barkai}},\ }\href {\doibase
  10.1103/PhysRevLett.101.058101} {\bibfield  {journal} {\bibinfo  {journal}
  {Phys. Rev. Lett.}\ }\textbf {\bibinfo {volume} {101}},\ \bibinfo {pages}
  {058101} (\bibinfo {year} {2008})}\BibitemShut {NoStop}%
\bibitem [{\citenamefont {Lubelski}\ \emph {et~al.}(2008)\citenamefont
  {Lubelski}, \citenamefont {Sokolov},\ and\ \citenamefont
  {Klafter}}]{LubelskiSK2008}%
  \BibitemOpen
  \bibfield  {author} {\bibinfo {author} {\bibfnamefont {A.}~\bibnamefont
  {Lubelski}}, \bibinfo {author} {\bibfnamefont {I.~M.}\ \bibnamefont
  {Sokolov}}, \ and\ \bibinfo {author} {\bibfnamefont {J.}~\bibnamefont
  {Klafter}},\ }\href {\doibase 10.1103/PhysRevLett.100.250602} {\bibfield
  {journal} {\bibinfo  {journal} {Phys. Rev. Lett.}\ }\textbf {\bibinfo
  {volume} {100}},\ \bibinfo {pages} {250602} (\bibinfo {year}
  {2008})}\BibitemShut {NoStop}%
\bibitem [{\citenamefont {Sokolov}(2008)}]{Sokolov2008}%
  \BibitemOpen
  \bibfield  {author} {\bibinfo {author} {\bibfnamefont {I.~M.}\ \bibnamefont
  {Sokolov}},\ }\href {\doibase 10.1103/Physics.1.8} {\bibfield  {journal}
  {\bibinfo  {journal} {Physics}\ }\textbf {\bibinfo {volume} {1}} (\bibinfo
  {year} {2008}),\ 10.1103/Physics.1.8}\BibitemShut {NoStop}%
\bibitem [{\citenamefont {Jeon}\ \emph {et~al.}(2011)\citenamefont {Jeon},
  \citenamefont {Tejedor}, \citenamefont {Burov}, \citenamefont {Barkai},
  \citenamefont {Selhuber-Unkel}, \citenamefont {Berg-S\o{}rensen},
  \citenamefont {Oddershede},\ and\ \citenamefont {Metzler}}]{Jeon2011}%
  \BibitemOpen
  \bibfield  {author} {\bibinfo {author} {\bibfnamefont {J.-H.}\ \bibnamefont
  {Jeon}}, \bibinfo {author} {\bibfnamefont {V.}~\bibnamefont {Tejedor}},
  \bibinfo {author} {\bibfnamefont {S.}~\bibnamefont {Burov}}, \bibinfo
  {author} {\bibfnamefont {E.}~\bibnamefont {Barkai}}, \bibinfo {author}
  {\bibfnamefont {C.}~\bibnamefont {Selhuber-Unkel}}, \bibinfo {author}
  {\bibfnamefont {K.}~\bibnamefont {Berg-S\o{}rensen}}, \bibinfo {author}
  {\bibfnamefont {L.}~\bibnamefont {Oddershede}}, \ and\ \bibinfo {author}
  {\bibfnamefont {R.}~\bibnamefont {Metzler}},\ }\href {\doibase
  10.1103/PhysRevLett.106.048103} {\bibfield  {journal} {\bibinfo  {journal}
  {Phys. Rev. Lett.}\ }\textbf {\bibinfo {volume} {106}},\ \bibinfo {pages}
  {048103} (\bibinfo {year} {2011})}\BibitemShut {NoStop}%
\bibitem [{\citenamefont {Khoury}\ \emph {et~al.}(2011)\citenamefont {Khoury},
  \citenamefont {Lacasta}, \citenamefont {Sancho},\ and\ \citenamefont
  {Lindenberg}}]{Khoury2011}%
  \BibitemOpen
  \bibfield  {author} {\bibinfo {author} {\bibfnamefont {M.}~\bibnamefont
  {Khoury}}, \bibinfo {author} {\bibfnamefont {A.~M.}\ \bibnamefont {Lacasta}},
  \bibinfo {author} {\bibfnamefont {J.~M.}\ \bibnamefont {Sancho}}, \ and\
  \bibinfo {author} {\bibfnamefont {K.}~\bibnamefont {Lindenberg}},\ }\href
  {\doibase 10.1103/PhysRevLett.106.090602} {\bibfield  {journal} {\bibinfo
  {journal} {Phys. Rev. Lett.}\ }\textbf {\bibinfo {volume} {106}},\ \bibinfo
  {pages} {090602} (\bibinfo {year} {2011})}\BibitemShut {NoStop}%
\bibitem [{\citenamefont {Kusumi}\ \emph {et~al.}(2012)\citenamefont {Kusumi},
  \citenamefont {Fujiwara}, \citenamefont {Chadda}, \citenamefont {Xie},
  \citenamefont {Tsunoyama}, \citenamefont {Kalay}, \citenamefont {Kasai},\
  and\ \citenamefont {Suzuki}}]{Kusumi2012}%
  \BibitemOpen
  \bibfield  {author} {\bibinfo {author} {\bibfnamefont {A.}~\bibnamefont
  {Kusumi}}, \bibinfo {author} {\bibfnamefont {T.~K.}\ \bibnamefont
  {Fujiwara}}, \bibinfo {author} {\bibfnamefont {R.}~\bibnamefont {Chadda}},
  \bibinfo {author} {\bibfnamefont {M.}~\bibnamefont {Xie}}, \bibinfo {author}
  {\bibfnamefont {T.~A.}\ \bibnamefont {Tsunoyama}}, \bibinfo {author}
  {\bibfnamefont {Z.}~\bibnamefont {Kalay}}, \bibinfo {author} {\bibfnamefont
  {R.~S.}\ \bibnamefont {Kasai}}, \ and\ \bibinfo {author} {\bibfnamefont
  {K.~G.}\ \bibnamefont {Suzuki}},\ }\href {\doibase
  10.1146/annurev-cellbio-100809-151736} {\bibfield  {journal} {\bibinfo
  {journal} {Annu. Rev. Cell Dev. Biol.}\ }\textbf {\bibinfo {volume} {28}},\
  \bibinfo {pages} {215} (\bibinfo {year} {2012})}\BibitemShut {NoStop}%
\bibitem [{\citenamefont {Massignan}\ \emph {et~al.}(2014)\citenamefont
  {Massignan}, \citenamefont {Manzo}, \citenamefont {Torreno-Pina},
  \citenamefont {Garc\'{i}a-Parajo}, \citenamefont {Lewenstein},\ and\
  \citenamefont {Lapeyre}}]{Massignan2014}%
  \BibitemOpen
  \bibfield  {author} {\bibinfo {author} {\bibfnamefont {P.}~\bibnamefont
  {Massignan}}, \bibinfo {author} {\bibfnamefont {C.}~\bibnamefont {Manzo}},
  \bibinfo {author} {\bibfnamefont {J.~A.}\ \bibnamefont {Torreno-Pina}},
  \bibinfo {author} {\bibfnamefont {M.~F.}\ \bibnamefont {Garc\'{i}a-Parajo}},
  \bibinfo {author} {\bibfnamefont {M.}~\bibnamefont {Lewenstein}}, \ and\
  \bibinfo {author} {\bibfnamefont {G.}~\bibnamefont {Lapeyre}},\ }\href
  {\doibase 10.1103/PhysRevLett.112.150603} {\bibfield  {journal} {\bibinfo
  {journal} {Phys. Rev. Lett.}\ }\textbf {\bibinfo {volume} {112}},\ \bibinfo
  {pages} {150603} (\bibinfo {year} {2014})}\BibitemShut {NoStop}%
\bibitem [{\citenamefont {Dentz}\ and\ \citenamefont
  {Berkowitz}(2005)}]{Dentz2005}%
  \BibitemOpen
  \bibfield  {author} {\bibinfo {author} {\bibfnamefont {M.}~\bibnamefont
  {Dentz}}\ and\ \bibinfo {author} {\bibfnamefont {B.}~\bibnamefont
  {Berkowitz}},\ }\href {\doibase 10.1103/PhysRevE.72.031110} {\bibfield
  {journal} {\bibinfo  {journal} {Phys. Rev. E}\ }\textbf {\bibinfo {volume}
  {72}},\ \bibinfo {pages} {031110} (\bibinfo {year} {2005})}\BibitemShut
  {NoStop}%
\bibitem [{\citenamefont {Dentz}\ and\ \citenamefont
  {Bolster}(2010)}]{Dentz2010}%
  \BibitemOpen
  \bibfield  {author} {\bibinfo {author} {\bibfnamefont {M.}~\bibnamefont
  {Dentz}}\ and\ \bibinfo {author} {\bibfnamefont {D.}~\bibnamefont
  {Bolster}},\ }\href {\doibase 10.1103/PhysRevLett.105.244301} {\bibfield
  {journal} {\bibinfo  {journal} {Phys. Rev. Lett.}\ }\textbf {\bibinfo
  {volume} {105}},\ \bibinfo {pages} {244301} (\bibinfo {year}
  {2010})}\BibitemShut {NoStop}%
\bibitem [{\citenamefont {Kang}\ \emph {et~al.}(2011)\citenamefont {Kang},
  \citenamefont {Dentz},\ and\ \citenamefont {Juanes}}]{Kang2011}%
  \BibitemOpen
  \bibfield  {author} {\bibinfo {author} {\bibfnamefont {P.~K.}\ \bibnamefont
  {Kang}}, \bibinfo {author} {\bibfnamefont {M.}~\bibnamefont {Dentz}}, \ and\
  \bibinfo {author} {\bibfnamefont {R.}~\bibnamefont {Juanes}},\ }\href
  {\doibase 10.1103/PhysRevE.83.030101} {\bibfield  {journal} {\bibinfo
  {journal} {Phys. Rev. E}\ }\textbf {\bibinfo {volume} {83}},\ \bibinfo
  {pages} {030101} (\bibinfo {year} {2011})}\BibitemShut {NoStop}%
\bibitem [{\citenamefont {K\"{u}hn}\ \emph {et~al.}(2011)\citenamefont
  {K\"{u}hn}, \citenamefont {Ihalainen}, \citenamefont {Hyv\"{a}luoma},
  \citenamefont {Dross}, \citenamefont {Willman}, \citenamefont {Langowski},
  \citenamefont {Vihinen-Ranta},\ and\ \citenamefont {Timonen}}]{Kuehn2011}%
  \BibitemOpen
  \bibfield  {author} {\bibinfo {author} {\bibfnamefont {T.}~\bibnamefont
  {K\"{u}hn}}, \bibinfo {author} {\bibfnamefont {T.~O.}\ \bibnamefont
  {Ihalainen}}, \bibinfo {author} {\bibfnamefont {J.}~\bibnamefont
  {Hyv\"{a}luoma}}, \bibinfo {author} {\bibfnamefont {N.}~\bibnamefont
  {Dross}}, \bibinfo {author} {\bibfnamefont {S.~F.}\ \bibnamefont {Willman}},
  \bibinfo {author} {\bibfnamefont {J.}~\bibnamefont {Langowski}}, \bibinfo
  {author} {\bibfnamefont {M.}~\bibnamefont {Vihinen-Ranta}}, \ and\ \bibinfo
  {author} {\bibfnamefont {J.}~\bibnamefont {Timonen}},\ }\href {\doibase
  10.1371/journal.pone.0022962} {\bibfield  {journal} {\bibinfo  {journal}
  {PLoS ONE}\ }\textbf {\bibinfo {volume} {6}},\ \bibinfo {pages} {e22962}
  (\bibinfo {year} {2011})}\BibitemShut {NoStop}%
\bibitem [{\citenamefont {Manzo}\ \emph {et~al.}(2015)\citenamefont {Manzo},
  \citenamefont {Torreno-Pina}, \citenamefont {Massignan}, \citenamefont
  {Lapeyre}, \citenamefont {Lewenstein},\ and\ \citenamefont
  {Garcia~Parajo}}]{Manzo2015}%
  \BibitemOpen
  \bibfield  {author} {\bibinfo {author} {\bibfnamefont {C.}~\bibnamefont
  {Manzo}}, \bibinfo {author} {\bibfnamefont {J.~A.}\ \bibnamefont
  {Torreno-Pina}}, \bibinfo {author} {\bibfnamefont {P.}~\bibnamefont
  {Massignan}}, \bibinfo {author} {\bibfnamefont {G.~J.}\ \bibnamefont
  {Lapeyre}}, \bibinfo {author} {\bibfnamefont {M.}~\bibnamefont {Lewenstein}},
  \ and\ \bibinfo {author} {\bibfnamefont {M.~F.}\ \bibnamefont
  {Garcia~Parajo}},\ }\href {\doibase 10.1103/PhysRevX.5.011021} {\bibfield
  {journal} {\bibinfo  {journal} {Phys. Rev. X}\ }\textbf {\bibinfo {volume}
  {5}},\ \bibinfo {pages} {011021} (\bibinfo {year} {2015})}\BibitemShut
  {NoStop}%
\bibitem [{\citenamefont {Cherstvy}\ and\ \citenamefont
  {Metzler}(2013)}]{Cherstvy2013a}%
  \BibitemOpen
  \bibfield  {author} {\bibinfo {author} {\bibfnamefont {A.~G.}\ \bibnamefont
  {Cherstvy}}\ and\ \bibinfo {author} {\bibfnamefont {R.}~\bibnamefont
  {Metzler}},\ }\href {\doibase 10.1039/C3CP53056F} {\bibfield  {journal}
  {\bibinfo  {journal} {Phys. Chem. Chem. Phys.}\ }\textbf {\bibinfo {volume}
  {15}},\ \bibinfo {pages} {20220} (\bibinfo {year} {2013})}\BibitemShut
  {NoStop}%
\bibitem [{\citenamefont {Cherstvy}\ \emph {et~al.}(2014)\citenamefont
  {Cherstvy}, \citenamefont {Chechkin},\ and\ \citenamefont
  {Metzler}}]{Cherstvy2014}%
  \BibitemOpen
  \bibfield  {author} {\bibinfo {author} {\bibfnamefont {A.~G.}\ \bibnamefont
  {Cherstvy}}, \bibinfo {author} {\bibfnamefont {A.~V.}\ \bibnamefont
  {Chechkin}}, \ and\ \bibinfo {author} {\bibfnamefont {R.}~\bibnamefont
  {Metzler}},\ }\href {\doibase 10.1039/C3SM52846D} {\bibfield  {journal}
  {\bibinfo  {journal} {Soft Matter}\ }\textbf {\bibinfo {volume} {10}},\
  \bibinfo {pages} {1591} (\bibinfo {year} {2014})}\BibitemShut {NoStop}%
\bibitem [{\citenamefont {Ben-Avraham}\ and\ \citenamefont
  {Havlin}(1982)}]{BenAvraham1982}%
  \BibitemOpen
  \bibfield  {author} {\bibinfo {author} {\bibfnamefont {D.}~\bibnamefont
  {Ben-Avraham}}\ and\ \bibinfo {author} {\bibfnamefont {S.}~\bibnamefont
  {Havlin}},\ }\href {http://stacks.iop.org/0305-4470/15/i=12/a=007} {\bibfield
   {journal} {\bibinfo  {journal} {Journal of Physics A: Mathematical and
  General}\ }\textbf {\bibinfo {volume} {15}},\ \bibinfo {pages} {L691}
  (\bibinfo {year} {1982})}\BibitemShut {NoStop}%
\bibitem [{\citenamefont {Gefen}\ \emph {et~al.}(1983)\citenamefont {Gefen},
  \citenamefont {Aharony},\ and\ \citenamefont {Alexander}}]{Gefen1983}%
  \BibitemOpen
  \bibfield  {author} {\bibinfo {author} {\bibfnamefont {Y.}~\bibnamefont
  {Gefen}}, \bibinfo {author} {\bibfnamefont {A.}~\bibnamefont {Aharony}}, \
  and\ \bibinfo {author} {\bibfnamefont {S.}~\bibnamefont {Alexander}},\ }\href
  {\doibase 10.1103/PhysRevLett.50.77} {\bibfield  {journal} {\bibinfo
  {journal} {Phys. Rev. Lett.}\ }\textbf {\bibinfo {volume} {50}},\ \bibinfo
  {pages} {77} (\bibinfo {year} {1983})}\BibitemShut {NoStop}%
\bibitem [{Note1()}]{Note1}%
  \BibitemOpen
  \bibinfo {note} {$\protect \ensuremath \protect \operatorname {Ein}({s,z}) =
  \DOTSB \sum@ \slimits@ _{k=1}^\infty (-1)^{k+1} \protect \frac
  {z^k}{k!(k+s)}$}\BibitemShut {NoStop}%
\end{thebibliography}%

\end{document}